\documentclass[twocolumn]{aastex63}

\usepackage{amsmath}
\usepackage{multirow}
\usepackage{xspace}
\usepackage{enumitem}
\submitjournal{ApJL}

\shorttitle{Numerical Model of PSR J0030+0451}
\shortauthors{Chen et al.}
\graphicspath{{./}{figures/}}

\newcommand{\rhogj}{\rho_\mathrm{GJ}}
\newcommand{\RLC}{R_{LC}}
\newcommand{\rmax}{r_\mathrm{max}}

\newcommand{\psr}{PSR\,J0030+0451\xspace}
\newcommand{\psrs}{J0030\xspace}
\newcommand{\nicer}{\textit{NICER}\xspace}

\begin{document}

\title{A Numerical Model for the Multi-wavelength Lightcurves of PSR J0030+0451}

\correspondingauthor{Alexander Y. Chen}
\email{alexc@astro.princeton.edu}

\author[0000-0002-4738-1168]{Alexander Y. Chen}
\affil{Department of Astrophysical Sciences, Princeton University, Princeton, NJ 08544, USA}

\author[0000-0002-0108-4774]{Yajie Yuan}
\affiliation{Center for Computational Astrophysics, Flatiron Institute, 162 Fifth Avenue, New York, NY 10010, USA}

\author[0000-0003-3902-3915]{Georgios Vasilopoulos}
\affiliation{Department of Astronomy, Yale University, PO Box 208101, New Haven, CT 06520-8101, USA}

\begin{abstract}
  Recent modeling of \emph{Neutron Star Interior Composition
    Explorer}(\nicer) observations of the millisecond pulsar \psr
  suggests that the magnetic field of the pulsar is non-dipolar. We
  construct a magnetic field configuration where foot points of the open
  field lines closely resemble the hotspot configuration from \nicer
  observations. Using this magnetic field as input, we perform
  force-free simulations of the magnetosphere of PSR J0030+0451, showing
  the three-dimensional structure of its plasma-filled
  magnetosphere. Making simple and physically motivated assumptions
  about the emitting regions, we are able to construct the
  multi-wavelength lightcurves that qualitatively agree with the
  corresponding observations. The agreement suggests that multipole
  magnetic structures are the key to modeling this type of pulsars, and
  can be used to constrain the magnetic inclination angle and the
  location of radio emission.
\end{abstract}


\section{Introduction} 
\label{sec:intro}

\psr (hereafter \psrs) is an isolated millisecond pulsar with a spin
period of $P\approx 4.87\,\mathrm{ms}$.  Recently the \nicer
collaboration mapped out the surface of \psrs with unprecedented detail
\citep{2019ApJ...887L..21R, 2019ApJ...887L..24M} by modelling the
  pulsed thermal X-ray emission. They revealed that in order to match
the observed X-ray lightcurve, the hotspots on the surface have to be
in the same rotational hemisphere, not antipodal as naively
expected. Furthermore, one of the hotspots needs to be elongated in the
azmuthal direction. Both features suggest that higher multipole
components are present near the stellar surface, channeling current and
energetic particles to heat the surface at these particular spots.

\psrs has been observed in all available wavelengths including radio,
X-rays, and gamma-rays \citep[see, e.g.][]{2009ApJ...699.1171A}. As
pointed out by \citet{2019ApJ...887L..23B}, there is an apparent
discrepancy between the observation angles from the \nicer fit and
previous radio/gamma-ray modelling \citep{2014ApJS..213....6J}. A model
taking into account the non-dipolar field line geometry near the
star may be key to settling this discrepancy.

Such theoretical effort already exists. \citet{2017ApJ...851..137G}
developed an analytic prescription to find the current carrying
regions on the stellar surface when the magnetic field is axisymmetric
around a magnetic axis.  \citet{2019MNRAS.490.1774L} applied this
prescription to map out hot regions on the stellar surface.  However the
\nicer hotspots clearly call for a non-axisymmetric configuration. Due
to the wealth of observational data for \psr, any magnetospheric model
should strive to not only explain the X-ray emission, but to reproduce the
multi-wavelength lightcurves simultaneously.

\begin{figure*}[t]
  \centering
  \plottwo{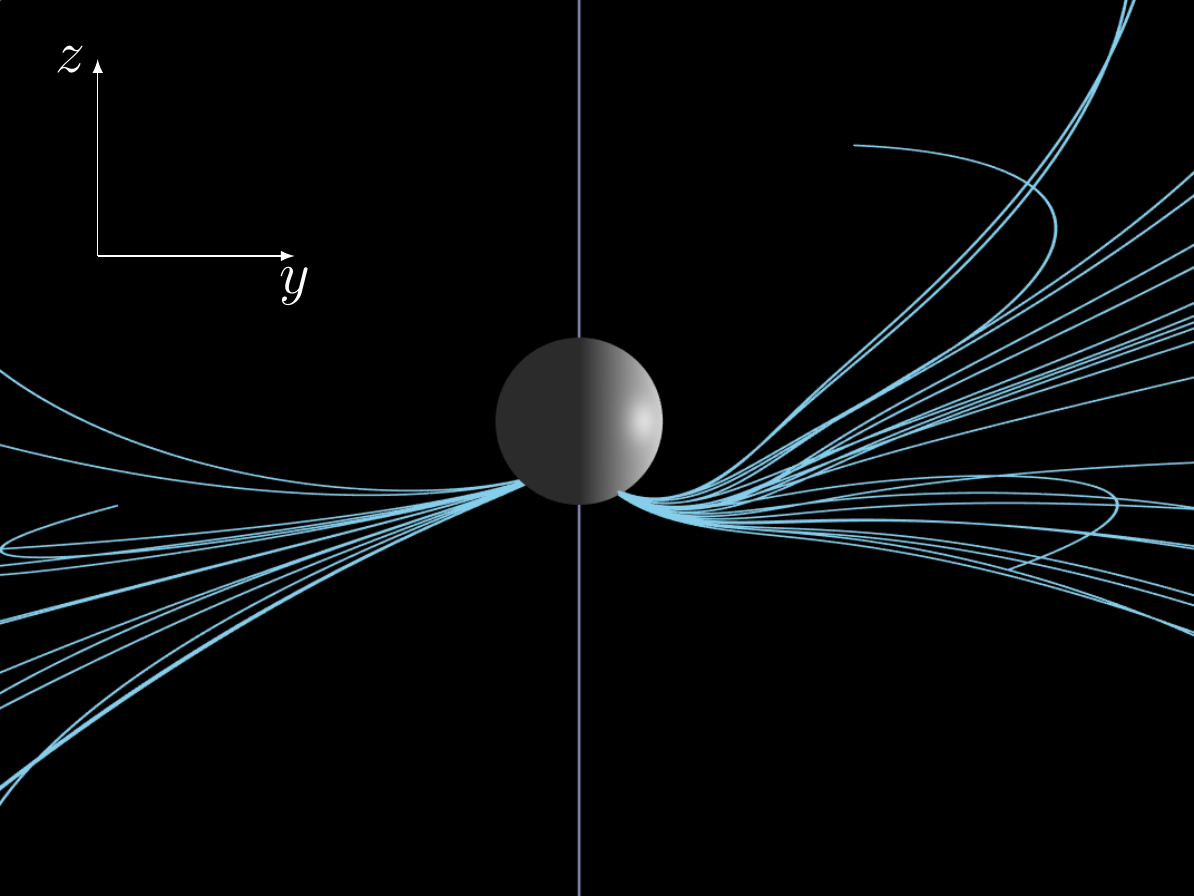}{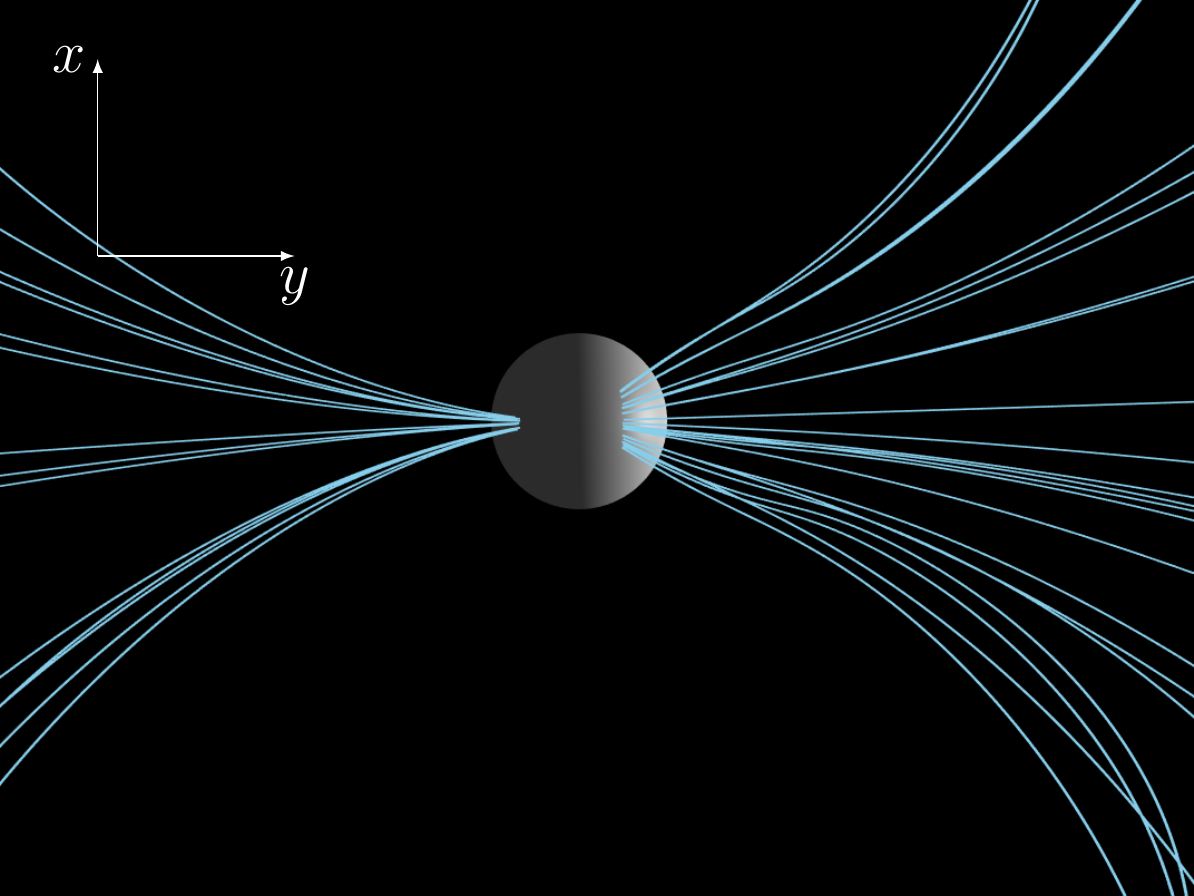}
  \caption{Vacuum magnetic field configuration obtained in our
    interactive tool, viewed from $x$ and $z$ axes. The parameters for
    this configuration are listed in section~\ref{sec:lightcurve}. The
    purple vertical line in the left panel is the rotation axis. Only
    field lines that extend beyond the light cylinder are drawn. A full
    version of this tool is hosted at \href{https://fizban007.github.io/PSRJ0030/field\_explorer.html}{https://fizban007.github.io/PSRJ0030/field\_explorer.html}.}
  \label{fig:vacuum-dipole}
\end{figure*}

In this Letter, we attempt to use the new results from \nicer
collaboration, together with recent insights from pulsar
theory, to construct a coherent emission model of \psrs in all observed
wavelengths. We first map out open field line regions using vacuum
dipole and quadrupole magnetic fields so that they resemble the \nicer
hotspots. Then we use force-free simulations to determine the global
magnetosphere structure and compute the numerical lightcurves.
\vspace{0.2cm}

\section{Vacuum Field Configuration} \label{sec:vacuum}

We expect that the hotspots on the surface of \psrs are
externally heated by plasma flow in the magnetosphere. In a
plasma-filled magnetosphere, electric current flows on open field lines,
and hits the star at the polar caps. A good starting point is
therefore to find a magnetic field configuration that has ``polar caps''
\footnote{Apparently the \nicer hotspots are no longer associated with
  magnetic poles, so "polar cap" is a misnomer. We will simply use this
  term to denote the collective foot points of open field lines.} with
shapes and positions similar to the reported hotspot patterns.

The hotspots found by the \nicer collaboration clearly require
multipole moments beyond the simple rotating dipole. Our first question
is whether they can be reproduced by simply using quadrupoles and not
higher multipole moments. To facilitate this, we developed a simple
interactive tool that integrates field lines originating from a pool of
seed points on the stellar surface, showing which field lines extend to
the light cylinder. We introduce the dipole moment vector $\mathbf{p}$
and the traceless symmetric quadrupole tensor $\mathbf{Q}$:
\begin{equation}
  \mathbf{Q} = 
  \begin{pmatrix}
    q_{11} & q_{12} & q_{13} \\
    q_{12} & q_{22} & q_{23} \\
    q_{13} & q_{23} & -q_{11} - q_{22}
  \end{pmatrix}.
\end{equation}
The most general static quadrupole field is defined to be:
\begin{equation}
  \mathbf{B}_q = -\nabla\left(\frac{\mathbf{r}\mathbf{Q}\mathbf{r}^T}{r^5}\right) = -\frac{2}{r^5}\mathbf{Q}\mathbf{r} + \frac{5}{r^7}(\mathbf{r}\mathbf{Q}\mathbf{r}^T)\mathbf{r}.
\end{equation}

Without loss of generality, we put the dipole moment in the $y$-$z$ plane, so
that $p_x = 0$. We also observe that in both
the hotspot configurations by \citet{2019ApJ...887L..21R} and
\citet{2019ApJ...887L..24M} the two regions are approximately spaced by
$180^\circ$ in $\phi$. Therefore, we attempt to keep everything symmetric
with respect to the $y$-$z$ plane, setting $q_{12} = q_{13} = 0$. We seek a
pair of polar caps that: 1. are both in the southern rotational
hemisphere; 2. one almost circular while the other significantly elongated
in the azmuthal direction.

We were initially unable to find vacuum configurations that satisfy the
above criteria with a combination of only dipolar and quadrupolar
fields. However, when we introduce an offset $z_\mathrm{offset}$ for the
center of the quadrupole component, we are able to find a range of
solutions with polar caps similar to the \nicer
results. Figure~\ref{fig:vacuum-dipole} shows an example of the vacuum
magnetic field obtained using our interactive tool.  There is, however,
a degeneracy in the inclination angle for the dipole component. We can
find such a polar cap configuration using a dipole inclination angle
between $65^\circ \sim 90^\circ$. We attempt to settle this degeneracy
using the gamma-ray lightcurves in the next section.

\section{Force-free Simulations and Dipole Gamma-ray Lightcurves} \label{sec:ffe}

\begin{figure*}[t]
  \centering
  \includegraphics[width=0.95\textwidth]{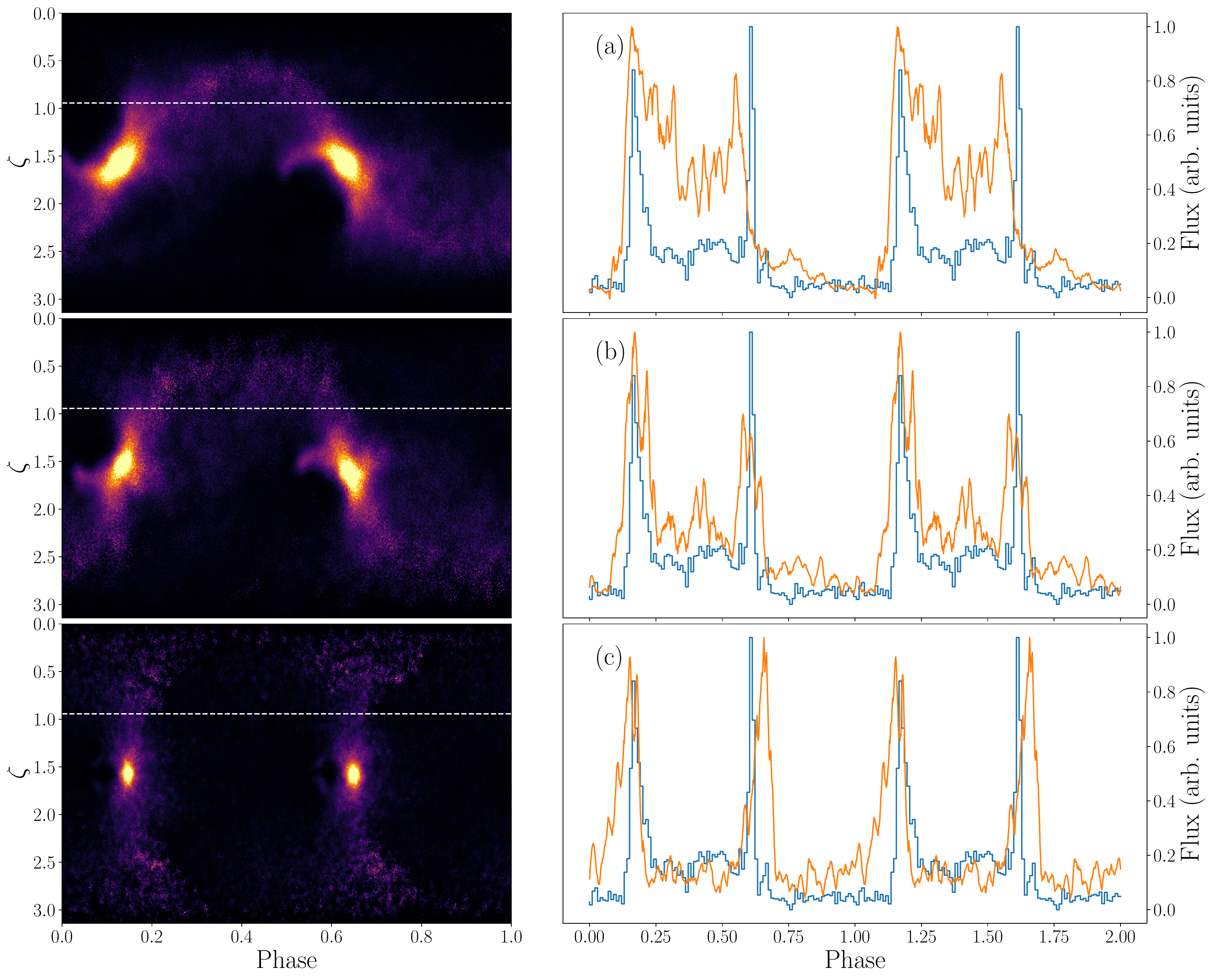}
  \caption{Comparison of the gamma-ray emission of 3 dipole rotators
    of different inclination angles. From top to bottom are (a)
    $\alpha = 60^{\circ}$, (b) $\alpha = 75^{\circ}$, and (c)
    $\alpha = 90^{\circ}$. Left side are the skymaps and right side
    are lightcurves sliced at $\zeta = 54^{\circ}$. Blue curves in
      the right panels are Fermi gamma-ray data of \psr at
      $>100\,\mathrm{MeV}$.}
  \label{fig:gamma-comparison}
\end{figure*}

PSR J0030+0451 is a strong pulsar that can easily produce $e^\pm$ pairs
in the current sheets near the light cylinder through $\gamma$-$\gamma$
collision \citep{2014ApJ...795L..22C, 2018ApJ...855...94P,
  2019ApJ...877...53H}. As a result, the magnetosphere is expected to be
plasma-filled and
well described by the force-free limit. The observed gamma-ray
emission is likely produced in these current sheets
\citep[e.g.,][]{2010ApJ...715.1282B,2015MNRAS.448..606C,2018ApJ...855...94P}. Since
the dipole field dominates over quadrupole near the light cylinder, we
first calculate the gamma-ray lightcurves from pure dipoles, and compare them with the observed one to constrain the dipole
inclination angle.

\begin{figure*}[t]
  \centering
  \begin{tabular}{c}
    \includegraphics[width=0.44\textwidth]{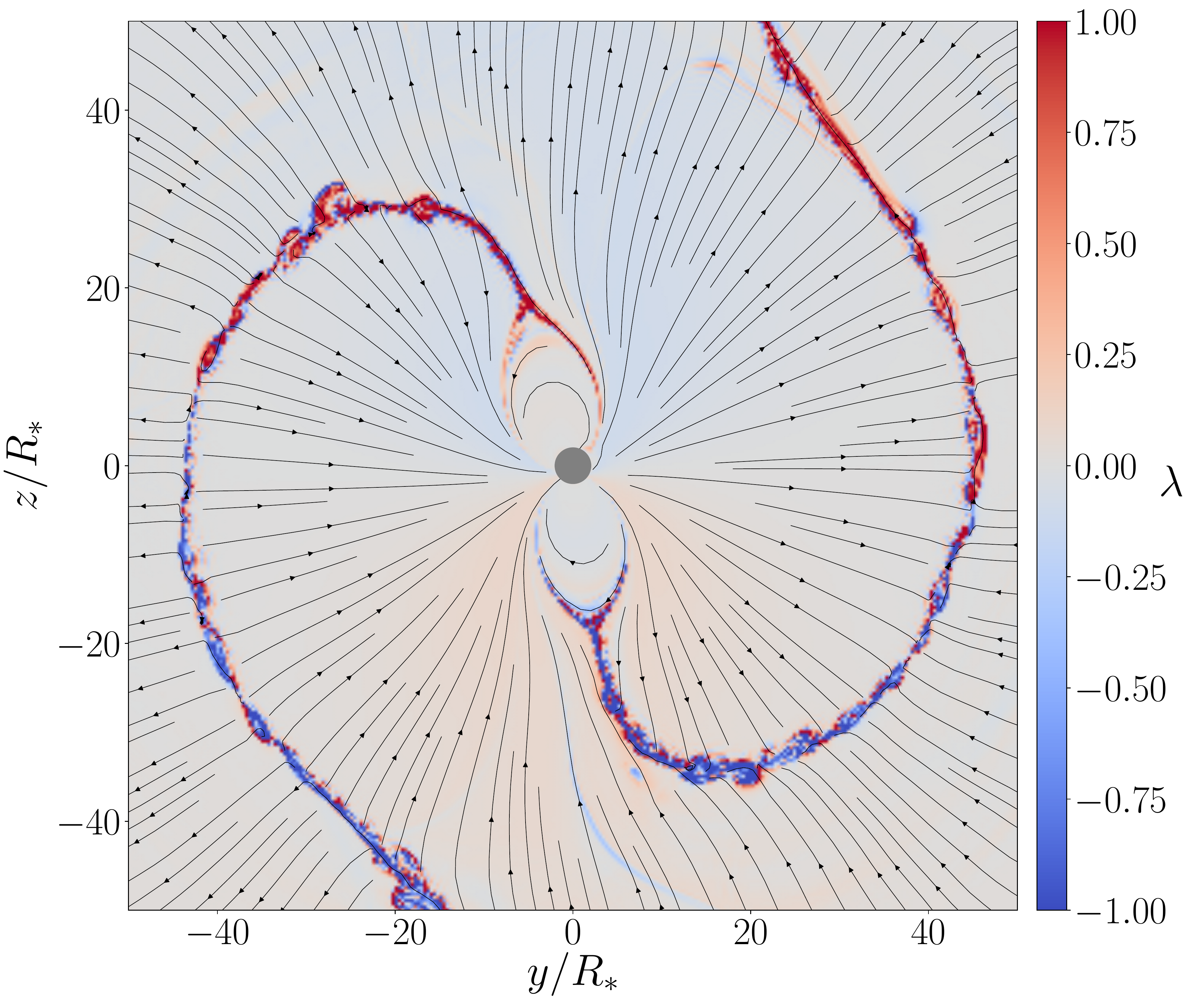}
    \includegraphics[width=0.54\textwidth]{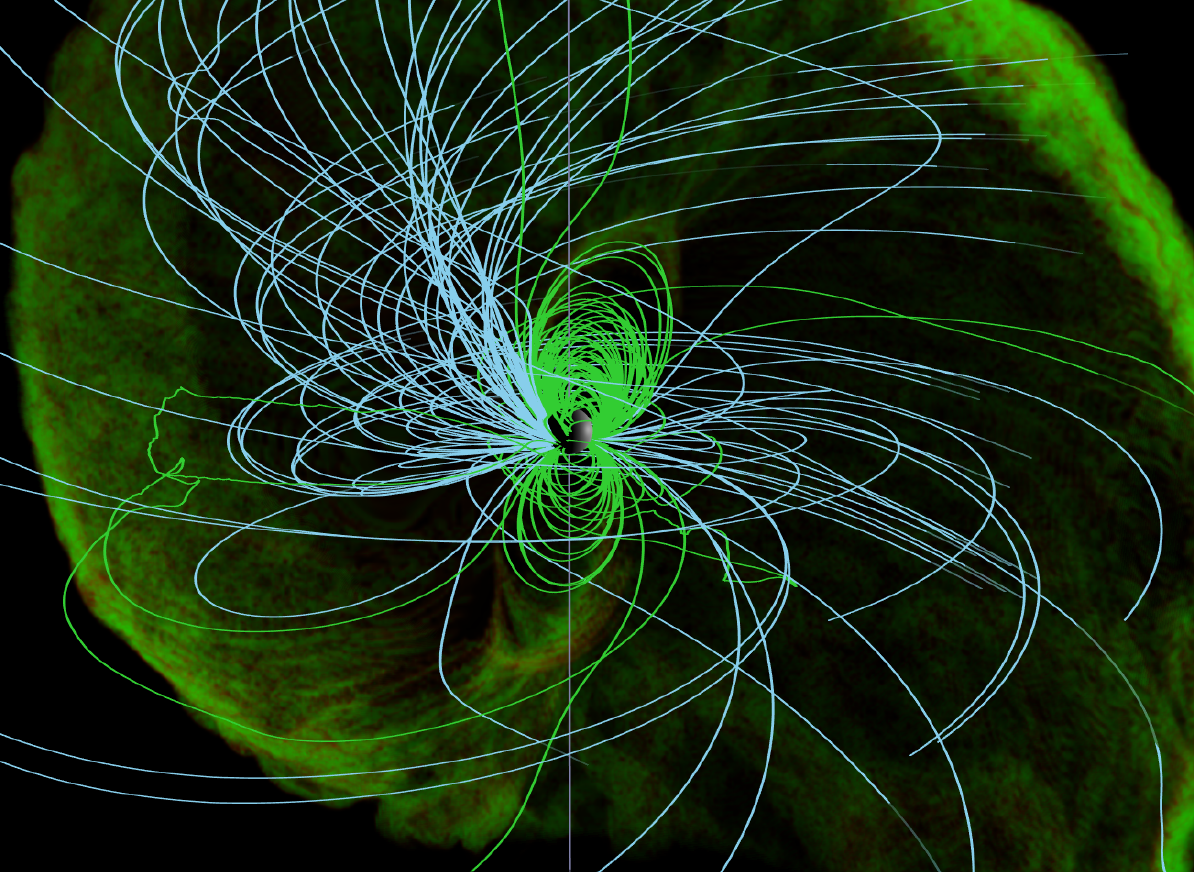} \\
    \includegraphics[width=0.98\textwidth]{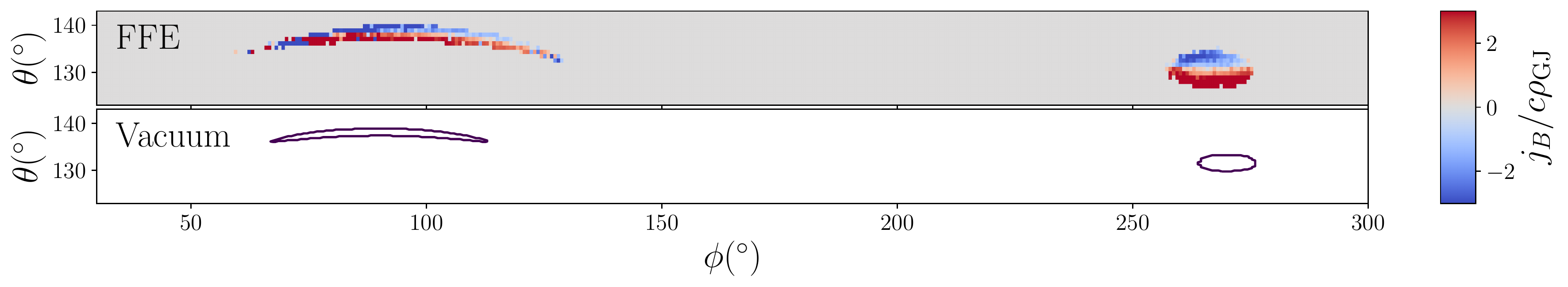}
  \end{tabular}
  \caption{Force-free magnetosphere of PSR J0030+0451 taken at
    $t=1.5T_\mathrm{rot}$. Upper left panel: a 2D slice of the quantity
    $\lambda$, with magnetic field as stream lines. Upper right panel: a
    3D rendering of the same snapshot. Green and cyan lines are closed
    and open field lines respectively. Volume rendering shows the 3D
    current sheet. Bottom panel top: polar cap distribution of
    $j_{B}/c\rhogj$ from the force-free simulation; bottom: vacuum polar
    cap of the configuration shown in Figure~\ref{fig:vacuum-dipole}. A
    fully interactive 3D render of the magnetosphere is hosted at
    \href{https://fizban007.github.io/PSRJ0030/ffe.html}{https://fizban007.github.io/PSRJ0030/ffe.html}.}
  \label{fig:lambda}
\end{figure*}

\begin{figure}[t]
  \centering
  \includegraphics[width=0.48\textwidth]{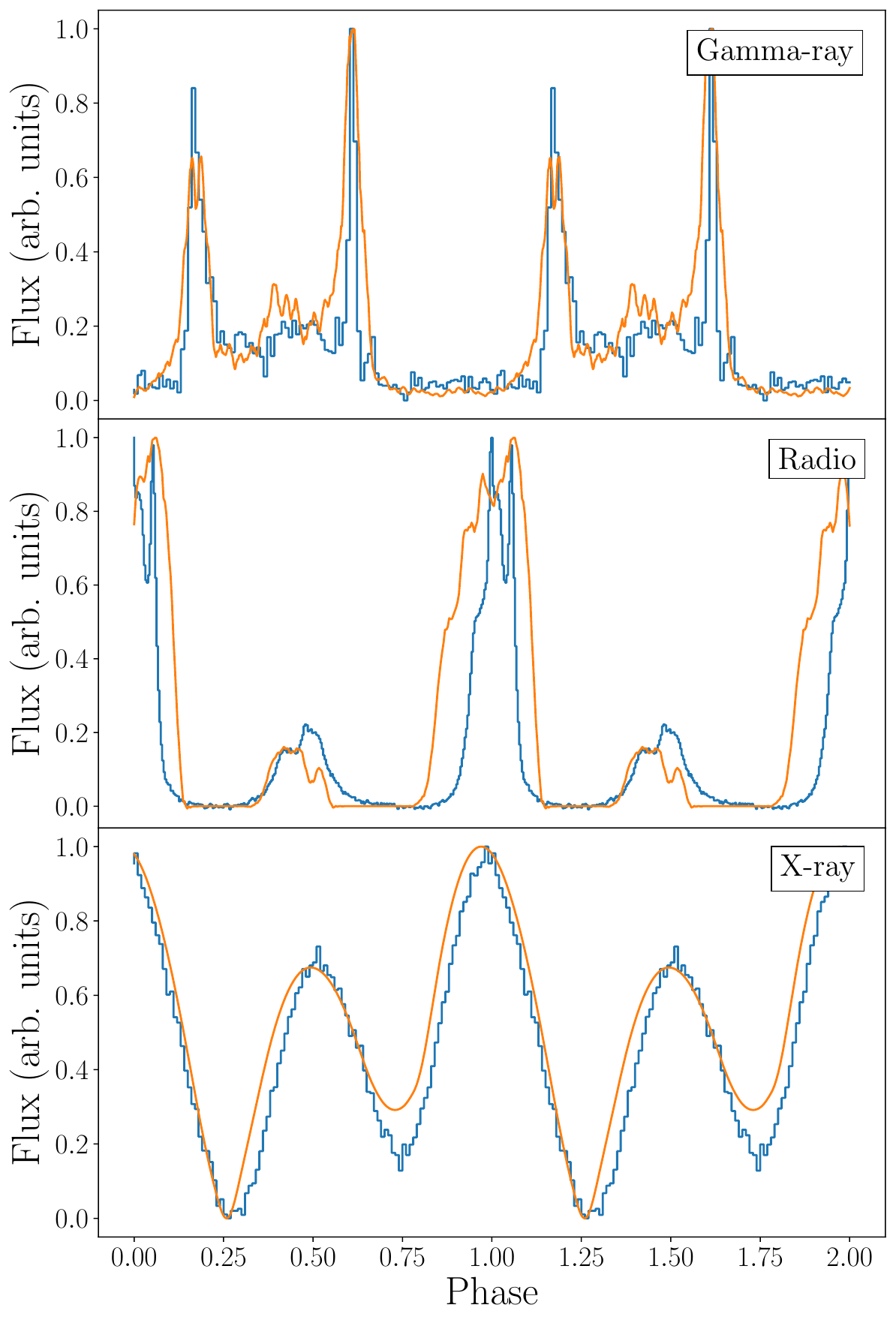}
  \caption{Multi-wavelength lightcurves for PSR J0030+0451. From top to
    bottom are gamma-rays ($>100\,\mathrm{MeV}$), radio
    ($1.4\,\mathrm{GHz}$), and X-rays (0.25-3.0 keV). For all curves the
    blue ones are from observations and orange from our numerical model. All
    curves are normalized to their maxima. Gamma-ray and radio data are taken from the second Fermi catalog \citep{2013ApJS..208...17A}, and X-ray data is from the \nicer dataset \citep{2019ApJ...887L..25B}.}
  \label{fig:light-curves}
\end{figure}

We use our own code \emph{Coffee} (COmputational Force FreE
Electrodynamics)\footnote{\href{https://github.com/fizban007/CoffeeGPU}{https://github.com/fizban007/CoffeeGPU}}
to solve the force-free equations:
\citep[e.g.,][]{1999astro.ph..2288G,2002luml.conf..381B}
\begin{align}
  \frac{\partial\mathbf{E}}{\partial t}&= \nabla\times\mathbf{B}-\mathbf{J},\\
  \frac{\partial\mathbf{B}}{\partial t}&=- \nabla\times\mathbf{E},\\
  \mathbf{J}&=\nabla\cdot\mathbf{E}\frac{\mathbf{E}\times\mathbf{B}}{B^2}+\frac{(\mathbf{B}\cdot\nabla\times\mathbf{B}-\mathbf{E}\cdot\nabla\times\mathbf{E})\mathbf{B}}{B^2},
\end{align}
with the constraints $\mathbf{E}\cdot\mathbf{B}=0$ and $E<B$ (we employ
Heaviside-Lorentz units and set $c=1$).  Our algorithm is similar to
\citet{2015PhRvL.115i5002E, 2016ApJ...817...89Z}: we use fourth-order
central finite difference stencils on a uniform Cartesian grid and a
five-stage fourth-order low storage Runge-Kutta scheme for time
evolution \citep{carpenter_fourth-order_1994}. We use hyperbolic
divergence cleaning \citep{2002JCoPh.175..645D} to damp any violations
of $\nabla\cdot\mathbf{B}=0$.  \footnote{Due to the higher
  order convergence of the scheme, even without divergence cleaning,
  $\nabla\cdot\mathbf{B}$ remains close to zero everywhere in the
  computational domain within the time range of our simulations.}  To
enforce the force-free condition, we explicitly remove any
$\mathbf{E}_{\parallel}$ by setting
$\mathbf{E}\to\mathbf{E}-(\mathbf{E}\cdot\mathbf{B})\mathbf{B}/B^2$ at
every time step. We apply standard sixth order Kreiss-Oliger numerical
dissipation to all hyperbolic variables to suppress high frequency noise
from truncation error \citep{kreiss_methods_1973}. To avoid stair
stepping at the pulsar surface, we force the fields to known values
inside the star with a smoothing kernel following
\citet{2006ApJ...648L..51S}. At the outer boundary, we implement an
absorbing layer to damp all outgoing electromagnetic waves
\citep[e.g.,][]{2015MNRAS.448..606C,2019MNRAS.487.4114Y}. The code is
parallelized and optimized to run on GPUs as well as CPUs with excellent
scaling.

The light cylinder radius $\RLC = c/\Omega_*$ of \psrs is approximately
$230\,\mathrm{km}$, or $\RLC/R_*\approx 20$. We use a Cartesian box of
size $6\RLC$ in each dimension, with resolution $1344^3$. However, this
resolution does not allow us to resolve $R_{*}$ well. Therefore, we set
the radius of the star to be at twice the real stellar radius
$r = 2R_*$, or $\RLC/r = 10$. This radius $r$ is resolved by 22 grid
points.

To find the gamma-ray lightcurve, we developed a method that focuses on
the emission from the current sheets. The main problem is that
since the polar caps are offset and irregular, it is difficult to use
the open volume coordinates $r_\mathrm{ov}$ defined by
\citet{2004ApJ...606.1125D} and invoked by
\citet{2010ApJ...715.1282B}. Instead, we look for current sheets in the
simulations directly. We define the quantity $\lambda$
\citep{2006ApJ...647L.119G}:
\begin{equation}
  \nabla\times\left( \mathbf{B} + \boldsymbol{\beta}_0 \times (\boldsymbol{\beta}_0 \times \mathbf{B}) \right) = \lambda \mathbf{B},
\end{equation}
where $\beta_0=\mathbf{\Omega}_{*}\times\mathbf{r}/c$.
$\lambda\mathbf{B}$ can be understood as the parallel force-free current
$\mathbf{j}_C$ in the corotating frame \citep[see,
e.g.][]{2010ApJ...715.1282B}. We identify regions where
$|\lambda| > 0.5$ as the current sheets (see Figure~\ref{fig:lambda} for
a map).  We place emitter particles in these cells between
$0.5\RLC < r < 1.5\RLC$. The motion of these particles consists of
parallel motion along the magnetic field lines as well as
$\mathbf{E}\times \mathbf{B}$ drift in the azmuthal direction:
\begin{equation}
  \mathbf{v} = x\mathbf{b} + \mathbf{v}_d,
\end{equation}
where $\mathbf{b} = \mathbf{B}/B$ is the direction of the magnetic
field, $\mathbf{v}_d = \mathbf{E}\times\mathbf{B}/B^2$ is the drift
velocity, and $x$ is a normalizing factor such that $|v|/c = 1$ and the
particle is moving outwards.  We allow for a small emission cone for
each particle of angular size $\delta\theta = 0.02$, and the actual
emission direction $\mathbf{e}$ is taken from a Gaussian distribution
centered around $\mathbf{v}$ with width $\delta\theta$.

To produce the skymap $(\phi, \zeta)$ where $\phi$ is the observation phase
and $\zeta$ is the observation angle, we subject the emission to the
usual time delay \citep[e.g.,][]{2010ApJ...715.1282B}:
\begin{equation}
  \label{eqn:delay}
  \phi = -\phi_e - \mathbf{e}\cdot\mathbf{r}/\RLC,
\end{equation}
where $\phi_e$ is the azimuthal direction of the emission direction $\mathbf{e}$.
We sum the contribution from each particle with a weight factor equal to
$j_C B$, which place the emphasis on the current sheet as well as taking
into account the local magnetic field.

To determine the dipole inclination $\alpha$, we ran a series of
simulations with pure dipole magnetic field and vary $\alpha$ from
$60^\circ$ to $90^\circ$. Figure~\ref{fig:gamma-comparison} shows the
skymaps and lightcurves from these simulations using the above
prescription. Since we have a separate constraint on viewing angle from
the \nicer observations, we could use the positions of the two gamma-ray
peaks as well as the amount of emission between peaks to determine the
inclination angle. The comparison seems to indicate that
$75^{\circ} < \alpha < 90^{\circ}$. We chose $\alpha = 80^{\circ}$ which
indeed gives a reasonable description for the observed lightcurve, as
can be seen in the top panel of Figure~\ref{fig:light-curves}.

\section{Numerical Model for PSR J0030+0451} \label{sec:lightcurve}

The final parameters we settle with are:
\begin{align}
  \mathbf{p} &= (p_x, p_y, p_z) = p_0(0, 0.985, 0.174), \\
  \mathbf{Q} &= p_0\begin{pmatrix}
    0.6 & 0 & 0 \\
    0 & -0.8 & -2 \\
    0 & -2 & 0.2
  \end{pmatrix}, \\
  z_\mathrm{offset} &= -0.4R_*.
\end{align}
This set of parameters corresponds to a dipole inclination angle of
$80^\circ$.
The quadrupole component is shifted and centered at
$(0, 0, z_\mathrm{offset})$.

Figure~\ref{fig:lambda} shows a global view of the force-free
magnetosphere. The force-free parameter $\lambda$ is indeed small
everywhere except in current sheets, providing a reliable way to
identify them. The closed field lines remain similar to the vacuum
configuration, whereas open field lines become mostly toroidal outside
the light cylinder. These features agree with force-free simulations
reported by \citet{2006ApJ...648L..51S}, and kinetic simulations by
\citet{2016MNRAS.457.2401C}, \citet{2018ApJ...855...94P}, and
\citet{2018ApJ...857...44K}. The global magnetosphere also remains
dipole-like, but the magnetospheric current is redirected to modified
polar caps due to the presence of higher multipole fields, in agreement
with the prediction by \citet{2017ApJ...851..137G}. Note that we did not
include general relativistic corrections to $\rhogj$, which will likely
reduce $\rhogj$ and enhance $j_B/c\rhogj$ \citep{2015ApJ...815L..19P,
  2020ApJ...889...69C}.

The bottom panel of Figure~\ref{fig:lambda} shows the polar caps from
both force-free and vacuum fields. The force-free polar caps are
obtained by integrating open field lines towards the star. Since the
simulation boundary condition is applied at $r < 2R_{*}$, we use the
vacuum field for the integration between
$R_{*}<r<2R_{*}$. $j_{B}/c\rhogj$ changes sign across both polar caps,
and is either negative or larger than unity. Note that although the
force-free polar caps closely resemble the vacuum ones, they are
  larger and slightly shifted. It is difficult to match the
force-free polar caps directly with \nicer results since a full fit
  using simulation results would take a prohibitive amount of
computational resources.

Figure~\ref{fig:light-curves} shows a comparison of different
lightcurves from our numerical model compared with the
observations. The peaks of the numerical lightcurves naturally
  line up with the data, without the need to individually shift each
  component. In the rest of this section, we discuss our method to
  compute the radio and X-ray lightcurves from the simulation results.

\subsection{Radio Emission}\label{sec:radio}

\begin{figure*}[t]
  \centering
  \includegraphics[width=0.95\textwidth]{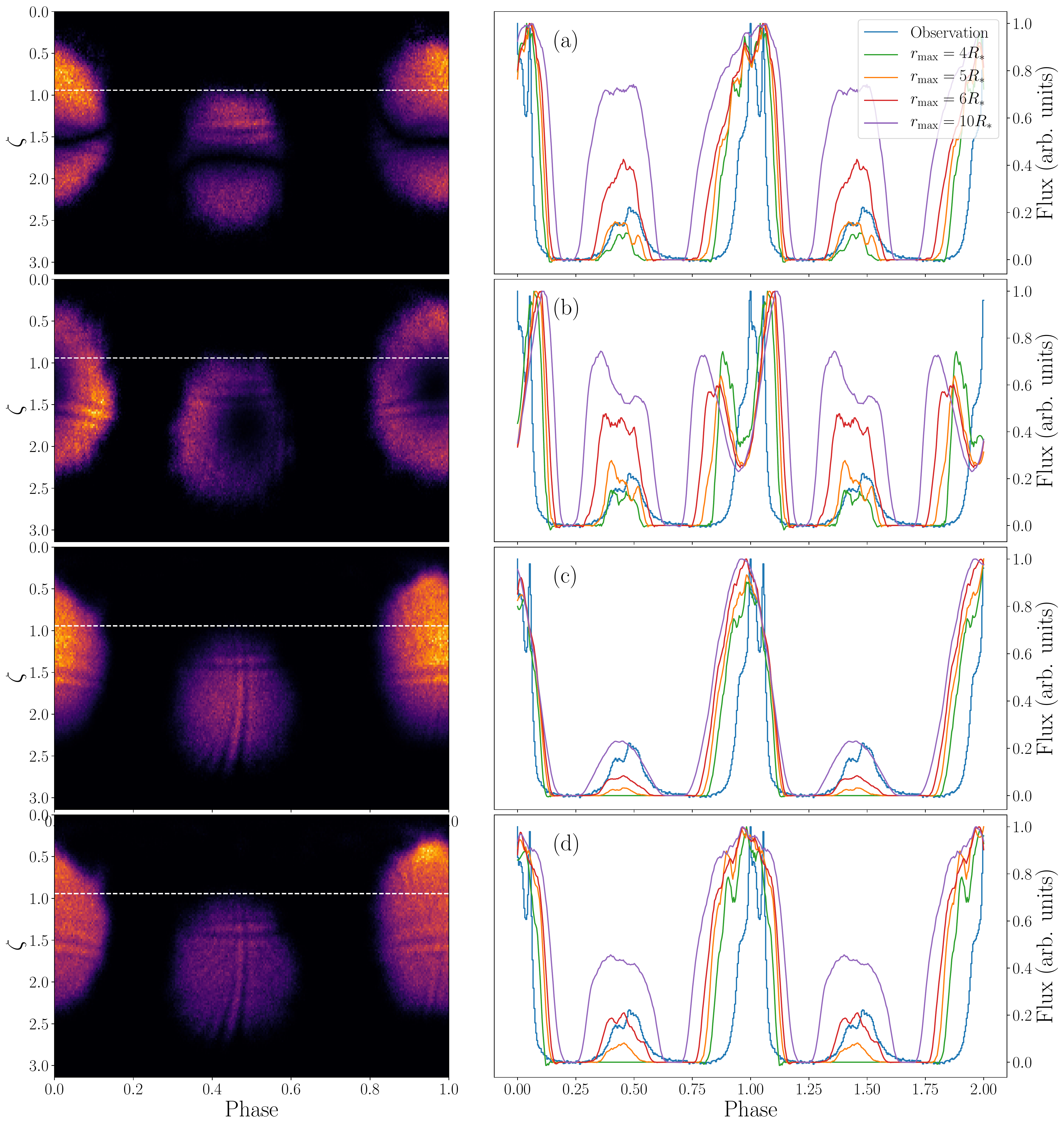}
  \caption{Comparison of the four weighting schemes for radio emission
    described in section~\ref{sec:radio}. From top to bottom are
    weighting schemes based on (a) the force-free current $\lambda$, (b)
    proximity to the edge of the polar cap, (c) proximity to the center
    of the polar cap, (d) uniform weight. The left panels are skymaps
    with $\rmax = 5R_{*}$ and right panels are the light curves seen by
    an observer at $\zeta = 54^\circ$. The different colors correspond
    to different $r_\mathrm{max}$. The stripe features in the
      skymaps are artifacts due to sampling a spherical emission region
      on a Cartesian grid.}
  \label{fig:radio-comparison}
\end{figure*}

To obtain the radio lightcurve, we adopt a prescription similar to the
gamma-rays, and use equation~\eqref{eqn:delay} to compute the arrival
phase of the signals emitted by test particles. Instead of identifying
the emitting region using $|\lambda|$, we assume all open field lines
between $r_\mathrm{min} = 2R_*$ and a variable $r_\mathrm{max}$ are
emitting. We sum up all the emission in this region and vary
$r_\mathrm{max}$ to try to determine the likely radii for radio
emission.

There still remains a significant degree of freedom in the weight we
assign to each emitting particle. We attempted several different weighting
schemes:

\begin{itemize}[leftmargin=*,topsep=2pt,itemsep=0.1pt]
\item Uniform emission weight.
\item Particle emission weight is proportional to $\lambda$.
\item For all the cells between $r$ and $r + R_*$ we define a mean
    emission direction by averaging the emission vector
    $\mathbf{e}$. The emission weight is proportional to
  $\sin^2\Delta\theta$ where $\Delta\theta$ is the angle between the
  emission direction and the mean direction. This ``ring-like'' scheme
  emphasizes the emission at the edge of the polar caps.
\item Similar as above, but weight proportional to
  $\cos^2\Delta\theta$. This ``center'' scheme emphasizes the geometric
  centers of the polar caps.
\end{itemize}
Figure~\ref{fig:radio-comparison} shows the results from the four
schemes above. In general, we always produce two radio peaks, one
large and one small, separated by approximately a half cycle. This is
consistent with the observed pattern. It can be seen that in general
larger $\rmax$ leads to higher interpulse. For each scheme, the relative
strengths of the two radio pulses single out an optimal
$r_\mathrm{max}$. The $\lambda$ scheme reflects the current structure of
the polar caps, showing split patterns on the skymap, a direct result of
the split polar caps shown in Figure~\ref{fig:lambda}. The ring-like
weighting scheme does indeed show a ring-like pattern on the skymap, and
tends to produce a double peak for the main radio peak. The center
scheme does not produce an appreciable interpulse between the main peaks
until $r_\mathrm{max} = 10R_*$. At this point both radio peaks are too
wide and arrive systematically earlier than the observed ones. The
$\lambda$ weighting is the most physically motivated, since the
magnetospheric current is what drives pair production
\citep{2008ApJ...683L..41B, 2013MNRAS.429...20T} and as a result, radio
emission. We find the $\lambda$ scheme with $r_\mathrm{max}=5$ is closest to the observations (see middle panel of Figure~\ref{fig:light-curves}).

The main radio peak in our best model is still wider than the observed
one. We believe this is because we assume the whole open field line
bundle is radio-emitting. Since this peak corresponds to the elongated
polar cap, it is conceivable that at the two corners of the polar cap
the parallel voltage is limited by the geometry, and pairs can only be
produced in the central region of the polar cap, resulting in a much
narrower radio-emitting region. This possibility needs to be
investigated further using self-consistent simulations. We focus
mostly on relatively large radii for radio emission, ignoring the
relativistic light bending effect which is important for the X-rays. This
effect may influence the contribution to the observed radio emission from
lower altitudes $r\lesssim 2R_{*}$.

\subsection{Hotspots and X-ray Lightcurve}

We obtain the force-free polar caps by tracing open field lines back to
$R_{*}$. The resulting polar caps are close to the vacuum polar
caps that we started with, which is a good consistency check. 
Both polar caps are split into
halves with different signs of current flowing, reminiscent of
near-orthogonal dipole rotators \citep[see,
e.g.][]{2013MNRAS.429...20T}. In both polar caps, $j_{B}/c\rhogj$ is
either negative (anti-GJ) or larger than unity (super-GJ), suggesting
that the whole polar cap should be active. As a first
approximation, we simply assume that both polar caps are heated
uniformly by the current flowing in the magnetosphere. 

We construct the X-ray lightcurve using the X-PSI package developed by
the Amsterdam group (Riley \& Watts 2019)
\footnote{\href{https://github.com/ThomasEdwardRiley/xpsi}{https://github.com/ThomasEdwardRiley/xpsi}}. The
current version of X-PSI lacks the ability to handle arbitrary-shaped
hotspots. Instead, we discretized the two hotspots on a $\theta$--$\phi$
grid, then put a small circular hotspot at the center of each occupied
grid point with uniform temperature $T = 1.3\times
10^{6}\,\mathrm{K}$. The shape of the polar caps can be found in the
bottom panel of Figure~\ref{fig:lambda} (we used the FFE
configuration). The stellar surface is taken to be cold
($T = 10^3\,\mathrm{K}$) and does not contribute to the \nicer observing
band. This ensemble of circular hotspots is then fed into X-PSI to
produce an ensemble of lightcurves. We sum all the lightcurves to
produce the one shown in the bottom panel of
Figure~\ref{fig:light-curves}. Again as a consistency check, the
lightcurve we obtained using this process is close to the
observations.

\section{Conclusion} \label{sec:conclusion}

We presented a numerical model that can reasonably reproduce the
lightcurves of PSR J0030+0451 at all observed frequencies including
radio, X-ray, and gamma-ray (Figure~\ref{fig:light-curves}). To achieve this, it is sufficient to
include only up to quadrupole magnetic field near the star with an
offset. We find that a dipole inclination angle of $\sim 80^\circ$
reproduces well the relative positions of the gamma-ray peaks.

The agreement of the new numerical model with observations strongly
suggests that electric current is indeed the driving factor for the
multi-wavelength emission in the pulsar magnetosphere. In addition, we
developed a method to simultaneously compute radio and gamma-ray
emission from a given magnetic field configuration using the force-free
current ratio $\lambda$. The radio emission height we obtained are
  not inconsistent with the phenomenological calculations by
  \citet{2003A&A...397..969K}, suggesting that our recipe can be
  potentially applied to other pulsars as well.

The simulations presented in this Letter are limited in resolution due
to our constraints on computation power. Future large-scale force-free
simulations should be able to better resolve the star, with stellar
surface at $R_*$ instead of $2R_*$. This will give a better
representation of the current distribution on the stellar surface and
better map to the configuration of hotspots. It could even be possible
to perform direct Particle-in-cell simulations of \psrs in the
foreseeable future, which will be able to pin-point the regions of
dissipation in the magnetosphere. PIC simulations will be able to
measure the amount of energy dissipated in the magnetosphere, and the
fraction of it which goes to heat the stellar surface, explaining the
origin and temperature of the hotspots on the star.

\acknowledgments We thank Anatoly Spitkovsky for helpful discussions and
comments on the manuscript. We also thank Jonathan Zrake for sharing the
force-free algorithm. The code \emph{Coffee} was developed initially
during the Princeton GPU hackathon in June 2019 by AC, YY, Hayk
Hakobyan, Patrick Crumley, and Vassilis Tsiolis. This work was supported
by NASA grant 80NSSC18K1099.  YY acknowledges support by a Flatiron
Research Fellowship at the Flatiron Institute, Simons Foundation.


\begin{thebibliography}{}
\expandafter\ifx\csname natexlab\endcsname\relax\def\natexlab#1{#1}\fi
\providecommand{\url}[1]{\href{#1}{#1}}
\providecommand{\dodoi}[1]{doi:~\href{http://doi.org/#1}{\nolinkurl{#1}}}
\providecommand{\doeprint}[1]{\href{http://ascl.net/#1}{\nolinkurl{http://ascl.net/#1}}}
\providecommand{\doarXiv}[1]{\href{https://arxiv.org/abs/#1}{\nolinkurl{https://arxiv.org/abs/#1}}}

\bibitem[{{Abdo} {et~al.}(2009){Abdo}, {Ackermann}, {Atwood}, {Axelsson},
  {Baldini}, {Ballet}, {Barbiellini}, {Bastieri}, {Battelino}, {Baughman},
  {Bechtol}, {Bellazzini}, {Berenji}, {Bloom}, {Bonamente}, {Borgland },
  {Bregeon}, {Brez}, {Brigida}, {Bruel}, {Burnett}, {Caliandro}, {Cameron},
  {Caraveo}, {Casandjian}, {Cecchi}, {Charles}, {Chekhtman}, {Cheung},
  {Chiang}, {Ciprini}, {Claus}, {Cognard}, {Cohen-Tanugi}, {Cominsky},
  {Conrad}, {Cutini}, {Dermer}, {de Angelis}, {de Palma}, {Digel}, {Dormody},
  {Silva}, {Drell}, {Dubois}, {Dumora}, {Farnier}, {Favuzzi}, {Focke},
  {Frailis}, {Fukazawa}, {Funk}, {Fusco}, {Gargano}, {Gasparrini}, {Gehrels},
  {Germani}, {Giebels}, {Giglietto}, {Giordano}, {Glanzman}, {Godfrey},
  {Grenier}, {Grondin}, {Grove}, {Guillemot}, {Guiriec}, {Hanabata}, {Harding},
  {Hayashida}, {Hays}, {Hughes}, {J{\'o}hannesson}, {Johnson}, {Johnson},
  {Johnson}, {Johnson}, {Kamae}, {Katagiri}, {Kataoka}, {Kawai}, {Kerr},
  {Kn{\"o}dlseder}, {Kocian}, {Komin}, {Kuehn}, {Kuss}, {Lande}, {Latronico},
  {Lee}, {Lemoine-Goumard}, {Longo}, {Loparco}, {Lott}, {Lovellette},
  {Lubrano}, {Madejski}, {Makeev}, {Marelli}, {Mazziotta}, {McConville},
  {McEnery}, {Meurer}, {Michelson}, {Mitthumsiri}, {Mizuno}, {Moiseev},
  {Monte}, {Monzani}, {Morselli}, {Moskalenko}, {Murgia}, {Nolan}, {Nuss},
  {Ohsugi}, {Omodei}, {Orlando}, {Ormes}, {Pancrazi}, {Paneque}, {Panetta},
  {Parent}, {Pepe}, {Pesce-Rollins}, {Piron}, {Porter}, {Rain{\`o}}, {Rando},
  {Razzano}, {Reimer}, {Reimer}, {Reposeur}, {Ritz}, {Rochester}, {Rodriguez},
  {Romani}, {Ryde}, {Sadrozinski}, {Sanchez}, {Sander}, {Parkinson},
  {Sgr{\`o}}, {Siskind}, {Smith}, {Smith}, {Spandre}, {Spinelli}, {Starck},
  {Strickman}, {Suson}, {Tajima}, {Takahashi}, {Tanaka}, {Thayer}, {Thayer},
  {Theureau}, {Thompson}, {Tibaldo}, {Torres}, {Tosti}, {Tramacere},
  {Uchiyama}, {Usher}, {Van Etten}, {Vilchez}, {Vitale}, {Waite}, {Watters},
  {Webb}, {Wood}, {Ylinen}, \& {Ziegler}}]{2009ApJ...699.1171A}
{Abdo}, A.~A., {Ackermann}, M., {Atwood}, W.~B., {et~al.} 2009, \apj, 699,
  1171, \dodoi{10.1088/0004-637X/699/2/1171}

\bibitem[{{Abdo} {et~al.}(2013){Abdo}, {Ajello}, {Allafort}, {Baldini},
  {Ballet}, {Barbiellini}, {Baring}, {Bastieri}, {Belfiore}, {Bellazzini},
  {Bhattacharyya}, {Bissaldi}, {Bloom}, {Bonamente}, {Bottacini}, {Brandt},
  {Bregeon}, {Brigida}, {Bruel}, {Buehler}, {Burgay}, {Burnett}, {Busetto},
  {Buson}, {Caliandro}, {Cameron}, {Camilo}, {Caraveo}, {Casandjian}, {Cecchi},
  {{\c{C}}elik}, {Charles}, {Chaty}, {Chaves}, {Chekhtman}, {Chen}, {Chiang},
  {Chiaro}, {Ciprini}, {Claus}, {Cognard}, {Cohen-Tanugi}, {Cominsky},
  {Conrad}, {Cutini}, {D'Ammando}, {de Angelis}, {DeCesar}, {De Luca}, {den
  Hartog}, {de Palma}, {Dermer}, {Desvignes}, {Digel}, {Di Venere}, {Drell},
  {Drlica-Wagner}, {Dubois}, {Dumora}, {Espinoza}, {Falletti}, {Favuzzi},
  {Ferrara}, {Focke}, {Franckowiak}, {Freire}, {Funk}, {Fusco}, {Gargano},
  {Gasparrini}, {Germani}, {Giglietto}, {Giommi}, {Giordano}, {Giroletti},
  {Glanzman}, {Godfrey}, {Gotthelf}, {Grenier}, {Grondin}, {Grove},
  {Guillemot}, {Guiriec}, {Hadasch}, {Hanabata}, {Harding}, {Hayashida},
  {Hays}, {Hessels}, {Hewitt}, {Hill}, {Horan}, {Hou}, {Hughes}, {Jackson},
  {Janssen}, {Jogler}, {J{\'o}hannesson}, {Johnson}, {Johnson}, {Johnson},
  {Johnson}, {Johnston}, {Kamae}, {Kataoka}, {Keith}, {Kerr}, {Kn{\"o}dlseder},
  {Kramer}, {Kuss}, {Lande}, {Larsson}, {Latronico}, {Lemoine-Goumard},
  {Longo}, {Loparco}, {Lovellette}, {Lubrano}, {Lyne}, {Manchester}, {Marelli},
  {Massaro}, {Mayer}, {Mazziotta}, {McEnery}, {McLaughlin}, {Mehault},
  {Michelson}, {Mignani}, {Mitthumsiri}, {Mizuno}, {Moiseev}, {Monzani},
  {Morselli}, {Moskalenko}, {Murgia}, {Nakamori}, {Nemmen}, {Nuss}, {Ohno},
  {Ohsugi}, {Orienti}, {Orlando}, {Ormes}, {Paneque}, {Panetta}, {Parent},
  {Perkins}, {Pesce-Rollins}, {Pierbattista}, {Piron}, {Pivato}, {Pletsch},
  {Porter}, {Possenti}, {Rain{\`o}}, {Rando}, {Ransom}, {Ray}, {Razzano},
  {Rea}, {Reimer}, {Reimer}, {Renault}, {Reposeur}, {Ritz}, {Romani}, {Roth},
  {Rousseau}, {Roy}, {Ruan}, {Sartori}, {Saz Parkinson}, {Scargle}, {Schulz},
  {Sgr{\`o}}, {Shannon}, {Siskind}, {Smith}, {Spandre}, {Spinelli}, {Stappers},
  {Strong}, {Suson}, {Takahashi}, {Thayer}, {Thayer}, {Theureau}, {Thompson},
  {Thorsett}, {Tibaldo}, {Tibolla}, {Tinivella}, {Torres}, {Tosti}, {Troja},
  {Uchiyama}, {Usher}, {Vandenbroucke}, {Vasileiou}, {Venter}, {Vianello},
  {Vitale}, {Wang}, {Weltevrede}, {Winer}, {Wolff}, {Wood}, {Wood}, {Wood}, \&
  {Yang}}]{2013ApJS..208...17A}
{Abdo}, A.~A., {Ajello}, M., {Allafort}, A., {et~al.} 2013, \apjs, 208, 17,
  \dodoi{10.1088/0067-0049/208/2/17}

\bibitem[{{Bai} \& {Spitkovsky}(2010)}]{2010ApJ...715.1282B}
{Bai}, X.-N., \& {Spitkovsky}, A. 2010, \apj, 715, 1282,
  \dodoi{10.1088/0004-637X/715/2/1282}

\bibitem[{{Beloborodov}(2008)}]{2008ApJ...683L..41B}
{Beloborodov}, A.~M. 2008, \apjl, 683, L41, \dodoi{10.1086/590079}

\bibitem[{{Bilous} {et~al.}(2019){Bilous}, {Watts}, {Harding}, {Riley},
  {Arzoumanian}, {Bogdanov}, {Gendreau}, {Ray}, {Guillot}, {Ho}, \&
  {Chakrabarty}}]{2019ApJ...887L..23B}
{Bilous}, A.~V., {Watts}, A.~L., {Harding}, A.~K., {et~al.} 2019, \apjl, 887,
  L23, \dodoi{10.3847/2041-8213/ab53e7}

\bibitem[{{Blandford}(2002)}]{2002luml.conf..381B}
{Blandford}, R.~D. 2002, in Lighthouses of the Universe: The Most Luminous
  Celestial Objects and Their Use for Cosmology, ed. M.~{Gilfanov},
  R.~{Sunyeav}, \& E.~{Churazov}, 381, \dodoi{10.1007/10856495_59}

\bibitem[{{Bogdanov} {et~al.}(2019){Bogdanov}, {Guillot}, {Ray}, {Wolff},
  {Chakrabarty}, {Ho}, {Kerr}, {Lamb}, {Lommen}, {Ludlam}, {Milburn},
  {Montano}, {Miller}, {Baub{\"o}ck}, {{\"O}zel}, {Psaltis}, {Remillard},
  {Riley}, {Steiner}, {Strohmayer}, {Watts}, {Wood}, {Zeldes}, {Enoto},
  {Okajima}, {Kellogg}, {Baker}, {Markwardt}, {Arzoumanian}, \&
  {Gendreau}}]{2019ApJ...887L..25B}
{Bogdanov}, S., {Guillot}, S., {Ray}, P.~S., {et~al.} 2019, \apjl, 887, L25,
  \dodoi{10.3847/2041-8213/ab53eb}

\bibitem[{Carpenter \& Kennedy(1994)}]{carpenter_fourth-order_1994}
Carpenter, M. H.~K., \& Kennedy, C.~A. 1994, Fourth-order {2N}-storage
  {Runge}-{Kutta} schemes, Technical {Report} NASA-TM-109112, NAS 1.15:109112,
  NASA Langley Research Center; Hampton, VA, United States.
\newblock \url{https://ntrs.nasa.gov/search.jsp?R=19940028444}

\bibitem[{{Cerutti} {et~al.}(2015){Cerutti}, {Philippov}, {Parfrey}, \&
  {Spitkovsky}}]{2015MNRAS.448..606C}
{Cerutti}, B., {Philippov}, A., {Parfrey}, K., \& {Spitkovsky}, A. 2015,
  \mnras, 448, 606, \dodoi{10.1093/mnras/stv042}

\bibitem[{{Cerutti} {et~al.}(2016){Cerutti}, {Philippov}, \&
  {Spitkovsky}}]{2016MNRAS.457.2401C}
{Cerutti}, B., {Philippov}, A.~A., \& {Spitkovsky}, A. 2016, \mnras, 457, 2401,
  \dodoi{10.1093/mnras/stw124}

\bibitem[{{Chen} \& {Beloborodov}(2014)}]{2014ApJ...795L..22C}
{Chen}, A.~Y., \& {Beloborodov}, A.~M. 2014, \apjl, 795, L22,
  \dodoi{10.1088/2041-8205/795/1/L22}

\bibitem[{{Chen} {et~al.}(2020){Chen}, {Cruz}, \&
  {Spitkovsky}}]{2020ApJ...889...69C}
{Chen}, A.~Y., {Cruz}, F., \& {Spitkovsky}, A. 2020, \apj, 889, 69,
  \dodoi{10.3847/1538-4357/ab5c20}

\bibitem[{{Dedner} {et~al.}(2002){Dedner}, {Kemm}, {Kr{\"o}ner}, {Munz},
  {Schnitzer}, \& {Wesenberg}}]{2002JCoPh.175..645D}
{Dedner}, A., {Kemm}, F., {Kr{\"o}ner}, D., {et~al.} 2002, Journal of
  Computational Physics, 175, 645, \dodoi{10.1006/jcph.2001.6961}

\bibitem[{{Dyks} {et~al.}(2004){Dyks}, {Harding}, \&
  {Rudak}}]{2004ApJ...606.1125D}
{Dyks}, J., {Harding}, A.~K., \& {Rudak}, B. 2004, \apj, 606, 1125,
  \dodoi{10.1086/383121}

\bibitem[{{East} {et~al.}(2015){East}, {Zrake}, {Yuan}, \&
  {Blandford}}]{2015PhRvL.115i5002E}
{East}, W.~E., {Zrake}, J., {Yuan}, Y., \& {Blandford}, R.~D. 2015, \prl, 115,
  095002, \dodoi{10.1103/PhysRevLett.115.095002}

\bibitem[{{Gralla} {et~al.}(2017){Gralla}, {Lupsasca}, \&
  {Philippov}}]{2017ApJ...851..137G}
{Gralla}, S.~E., {Lupsasca}, A., \& {Philippov}, A. 2017, \apj, 851, 137,
  \dodoi{10.3847/1538-4357/aa978d}

\bibitem[{{Gruzinov}(1999)}]{1999astro.ph..2288G}
{Gruzinov}, A. 1999, ArXiv e-prints, astro.
\newblock \doarXiv{astro-ph/9902288}

\bibitem[{{Gruzinov}(2006)}]{2006ApJ...647L.119G}
---. 2006, \apjl, 647, L119, \dodoi{10.1086/506590}

\bibitem[{{Hakobyan} {et~al.}(2019){Hakobyan}, {Philippov}, \&
  {Spitkovsky}}]{2019ApJ...877...53H}
{Hakobyan}, H., {Philippov}, A., \& {Spitkovsky}, A. 2019, \apj, 877, 53,
  \dodoi{10.3847/1538-4357/ab191b}

\bibitem[{{Johnson} {et~al.}(2014){Johnson}, {Venter}, {Harding}, {Guillemot},
  {Smith}, {Kramer}, {{\c{C}}elik}, {den Hartog}, {Ferrara}, {Hou}, {Lande}, \&
  {Ray}}]{2014ApJS..213....6J}
{Johnson}, T.~J., {Venter}, C., {Harding}, A.~K., {et~al.} 2014, \apjs, 213, 6,
  \dodoi{10.1088/0067-0049/213/1/6}

\bibitem[{{Kalapotharakos} {et~al.}(2018){Kalapotharakos}, {Brambilla},
  {Timokhin}, {Harding}, \& {Kazanas}}]{2018ApJ...857...44K}
{Kalapotharakos}, C., {Brambilla}, G., {Timokhin}, A., {Harding}, A.~K., \&
  {Kazanas}, D. 2018, \apj, 857, 44, \dodoi{10.3847/1538-4357/aab550}

\bibitem[{{Kijak} \& {Gil}(2003)}]{2003A&A...397..969K}
{Kijak}, J., \& {Gil}, J. 2003, \aap, 397, 969,
  \dodoi{10.1051/0004-6361:20021583}

\bibitem[{Kreiss \& Oliger(1973)}]{kreiss_methods_1973}
Kreiss, H.~O., \& Oliger, J. 1973, Methods for the approximate solution of time
  dependent problems, {GARP} publications series No.~10 (Geneva: Global
  Atmospheric Research Programme - WMO-ICSU Joint Organizing Committee)

\bibitem[{{Lockhart} {et~al.}(2019){Lockhart}, {Gralla}, {{\"O}zel}, \&
  {Psaltis}}]{2019MNRAS.490.1774L}
{Lockhart}, W., {Gralla}, S.~E., {{\"O}zel}, F., \& {Psaltis}, D. 2019, \mnras,
  490, 1774, \dodoi{10.1093/mnras/stz2524}

\bibitem[{{Miller} {et~al.}(2019){Miller}, {Lamb}, {Dittmann}, {Bogdanov},
  {Arzoumanian}, {Gendreau}, {Guillot}, {Harding}, {Ho}, {Lattimer}, {Ludlam},
  {Mahmoodifar}, {Morsink}, {Ray}, {Strohmayer}, {Wood}, {Enoto}, {Foster},
  {Okajima}, {Prigozhin}, \& {Soong}}]{2019ApJ...887L..24M}
{Miller}, M.~C., {Lamb}, F.~K., {Dittmann}, A.~J., {et~al.} 2019, \apjl, 887,
  L24, \dodoi{10.3847/2041-8213/ab50c5}

\bibitem[{{Philippov} {et~al.}(2015){Philippov}, {Cerutti}, {Tchekhovskoy}, \&
  {Spitkovsky}}]{2015ApJ...815L..19P}
{Philippov}, A.~A., {Cerutti}, B., {Tchekhovskoy}, A., \& {Spitkovsky}, A.
  2015, \apjl, 815, L19, \dodoi{10.1088/2041-8205/815/2/L19}

\bibitem[{{Philippov} \& {Spitkovsky}(2018)}]{2018ApJ...855...94P}
{Philippov}, A.~A., \& {Spitkovsky}, A. 2018, \apj, 855, 94,
  \dodoi{10.3847/1538-4357/aaabbc}

\bibitem[{{Riley} {et~al.}(2019){Riley}, {Watts}, {Bogdanov}, {Ray}, {Ludlam},
  {Guillot}, {Arzoumanian}, {Baker}, {Bilous}, {Chakrabarty}, {Gendreau},
  {Harding}, {Ho}, {Lattimer}, {Morsink}, \&
  {Strohmayer}}]{2019ApJ...887L..21R}
{Riley}, T.~E., {Watts}, A.~L., {Bogdanov}, S., {et~al.} 2019, \apjl, 887, L21,
  \dodoi{10.3847/2041-8213/ab481c}

\bibitem[{{Spitkovsky}(2006)}]{2006ApJ...648L..51S}
{Spitkovsky}, A. 2006, \apjl, 648, L51, \dodoi{10.1086/507518}

\bibitem[{{Timokhin} \& {Arons}(2013)}]{2013MNRAS.429...20T}
{Timokhin}, A.~N., \& {Arons}, J. 2013, \mnras, 429, 20,
  \dodoi{10.1093/mnras/sts298}

\bibitem[{{Yuan} {et~al.}(2019){Yuan}, {Spitkovsky}, {Blandford}, \&
  {Wilkins}}]{2019MNRAS.487.4114Y}
{Yuan}, Y., {Spitkovsky}, A., {Blandford}, R.~D., \& {Wilkins}, D.~R. 2019,
  \mnras, 487, 4114, \dodoi{10.1093/mnras/stz1599}

\bibitem[{{Zrake} \& {East}(2016)}]{2016ApJ...817...89Z}
{Zrake}, J., \& {East}, W.~E. 2016, \apj, 817, 89,
  \dodoi{10.3847/0004-637X/817/2/89}

\end{thebibliography}
\bibliographystyle{aasjournal}



\end{document}